\documentclass[aps,prl,reprint,superscriptaddress]{revtex4-1}
\usepackage{graphicx}
\usepackage{color,amsmath}

\begin{document}


\title{Generation and Detection of Spin Currents in Semiconductor Nanostructures with Strong Spin-Orbit Interaction}



\author{Fabrizio Nichele}
\altaffiliation[Present Address: ]{Center for Quantum Devices, Niels Bohr Institute, University of Copenhagen, 2100 Copenhagen, Denmark}
\email[email: ]{fnichele@phys.ethz.ch}
\homepage[homepage: ]{www.nanophys.ethz.ch}
\affiliation{Solid State Physics Laboratory, ETH Z\"{u}rich, 8093 Z\"{u}rich, Switzerland}

\author{Szymon Hennel}
\affiliation{Solid State Physics Laboratory, ETH Z\"{u}rich, 8093 Z\"{u}rich, Switzerland}

\author{Patrick Pietsch}
\affiliation{Solid State Physics Laboratory, ETH Z\"{u}rich, 8093 Z\"{u}rich, Switzerland}

\author{Werner Wegscheider}
\affiliation{Solid State Physics Laboratory, ETH Z\"{u}rich, 8093 Z\"{u}rich, Switzerland}

\author{Peter Stano}
\affiliation{RIKEN Center for Emergent Matter Science, 2-1 Hirosawa, Wako, Saitama 351-0198, Japan}
\affiliation{Institute of Physics, Slovak Academy of Sciences, Dubravska cesta 9, 84511 Bratislava, Slovakia}

\author{Philippe Jacquod}
\affiliation{HES-SO, Haute Ecole d'Ing\'enierie, 1950 Sion, Switzerland}

\author{Thomas Ihn}
\affiliation{Solid State Physics Laboratory, ETH Z\"{u}rich, 8093 Z\"{u}rich, Switzerland}

\author{Klaus Ensslin}
\affiliation{Solid State Physics Laboratory, ETH Z\"{u}rich, 8093 Z\"{u}rich, Switzerland}


\date{\today}

\begin{abstract}

Storing, transmitting, and manipulating information using the electron spin resides at the heart of spintronics. Fundamental for future spintronics applications is the ability to control spin currents in solid state systems. Among the different platforms proposed so far, semiconductors with strong spin-orbit interaction are especially attractive as they promise fast and scalable spin control with all-electrical protocols. Here we demonstrate both the generation and measurement of pure spin currents in semiconductor nanostructures. Generation is purely electrical and mediated by the spin dynamics in materials with a strong spin-orbit field. Measurement is accomplished using a spin-to-charge conversion technique, based on the magnetic field symmetry of easily measurable electrical quantities. Calibrating the spin-to-charge conversion via the conductance of a quantum point contact, we quantitatively measure the mesoscopic spin Hall effect in a multiterminal GaAs dot. We report spin currents of $174~\rm{pA}$, corresponding to a spin Hall angle of $34\%$.
\end{abstract}


\maketitle


The generation and detection of spin currents in nanostructures is the central challenge of semiconductor spintronics. On the one hand, spin injection cannot be easily achieved by coupling semiconductors to ferromagnets \cite{Schmidt2000} because of the lack of control over material interfaces \cite{Sharma2014}. On the other hand, magnetoelectric alternatives exploiting the celebrated spin Hall effect (SHE) \cite{Dyakonov1971,Jungwirth2012}, have delivered only qualitative measurement protocols in transport experiments \cite{Brune2010}.
Alternatively to all-electrical setups, spin polarizing the current through a quantum point contact (QPC) with a magnetic field allows a quantitative control over spin current generation and detection at the nanoscale \cite{Potok2002,Watson2003,Frolov2009}. The latter approach typically requires such high magnetic fields ($6-8$ Tesla) that, as a drawback, the desired magnetoelectric effects are either suppressed or totally altered.

This Letter reports two major advances of nanoscale semiconductor spintronics. Namely, we develop novel experimental methods to \textit{electrically generate} and \textit{quantitatively measure} spin currents in a two-dimensional semiconductor nanostructure. 

\begin{figure}
\includegraphics[width=\columnwidth]{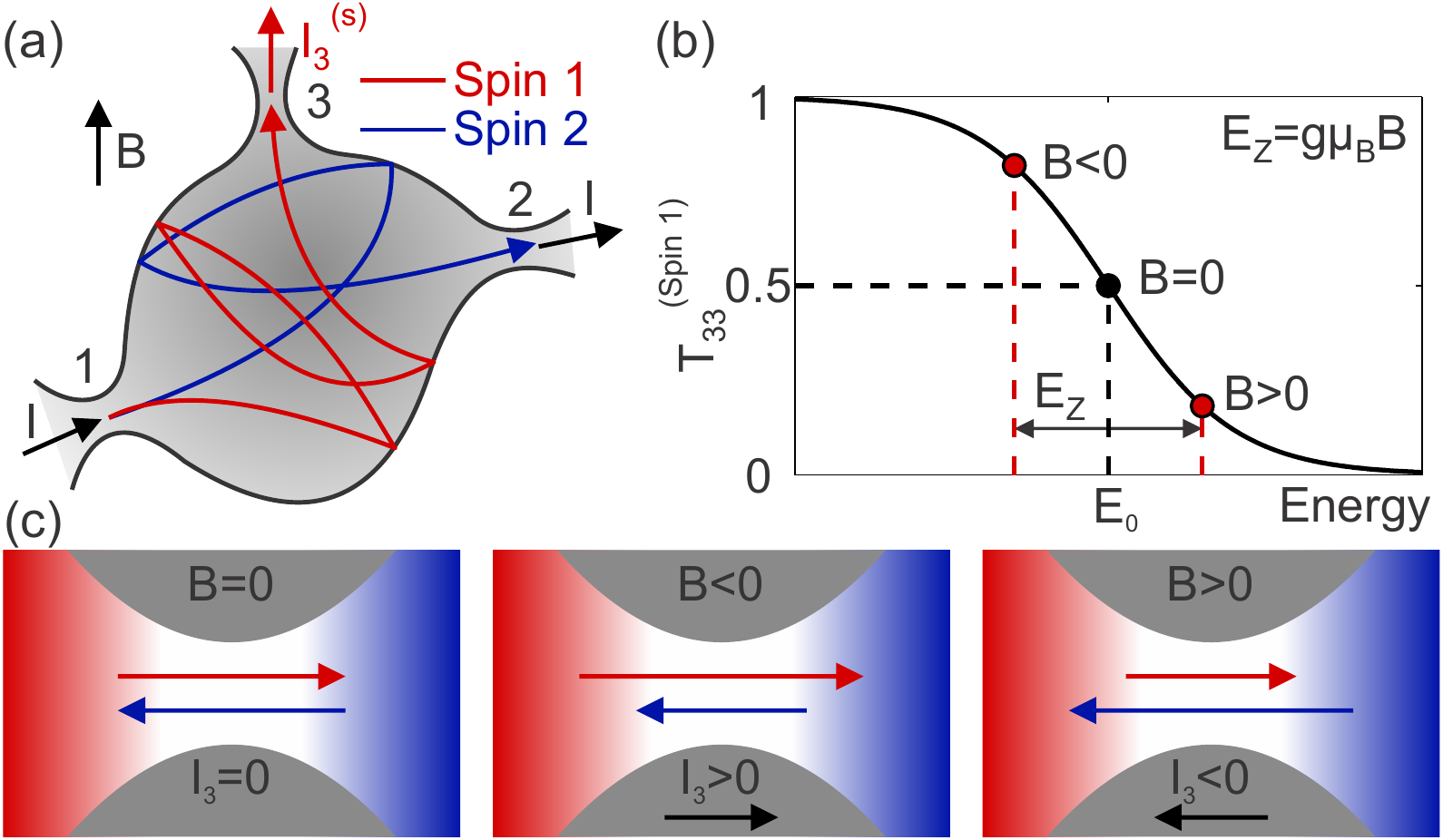}
\caption{(a) Schematic of the system used to generate spin currents. Charge currents are depicted as black lines, spin currents are depicted as red and blue lines. (b) Energy dependent, spin sensitive, transmission probability of $\rm{QPC_3}$. At zero field the transmission is tuned to $1/2$. A positive (negative) field suppresses (enhances) the transmission probability of one spin eigenstate. (c) Representation of spin and charge currents in $\rm{QPC_3}$ as a function of magnetic field. The net charge current in $\rm{QPC_3}$ varies with the magnetic field.}
\label{fig1}
\end{figure}

It is predicted that charge currents flowing through spin-orbit interaction (SOI)-coupled nanostructures are generically accompanied by spin currents, if the spin-orbit time is shorter than the electron dwell time~\cite{Ren2006,Bardarson2007,Nazarov2007,Krich2008}. This spin current generation mechanism is purely electrical and based on the mesoscopic SHE (MSHE) \cite{Ren2006,Bardarson2007}, where the electronic orbital dynamics in chaotic nanostructures cooperates with the SOI to make transport spin dependent. We will consider an open three-terminal quantum dot as represented in Fig.~\ref{fig1}(a), where each lead $i$ is a QPC carrying $N_i$ spin degenerate modes. Running a charge current $I$ between terminals $1$ and $2$, a spin current in all terminals, including $3$, is expected due to the MSHE.

For a weak SOI, the spin currents' amplitude fluctuates from sample to sample with zero average. For cavities with a strong SOI, geometric correlations between the spin and the orbital electronic dynamics lead to spin currents with large, predictable nonzero average values \cite{Adagideli2010}. In the latter case, the spin currents' amplitude is determined by the nanostructure geometry and the SOI strength. This particular mechanism renders spin currents robust against decoherence and allows us to differentiate them from mesoscopic fluctuations. This is essential for spintronics applications, where spin currents must be reproducible regardless of the microscopic details of the sample.

To detect and measure the spin currents described above, we employ the scheme of Ref.~\cite{Stano2011}, based on the magnetic field parity of the voltage behind a QPC. With reference to Fig.~\ref{fig1}(a), we are interested in the spin current $I_3^{(S)}$ emitted from $\rm{QPC_3}$. The energy dependent transmission probability of $\rm{QPC_3}$, $T_{33}^{(s)}$, is shown in Fig.~\ref{fig1}(b). At zero field, $\rm{QPC_3}$ is tuned to a conductance of $e^2/h$ by a suitable gate voltage, corresponding to a spin-independent transmission probability of one half. A weak in-plane magnetic field $B$ modifies the electrons' kinetic energy via Zeeman coupling, selectively increasing or decreasing the transmission probability according to the spin eigenstate and magnetic field sign. For simplicity, we assume $I^{(S)}_3$ to be a pure spin current at $B=0$, arising as two counterpropagating and inversely spin-polarized charge currents of equal magnitude, as schematically shown in Fig.~\ref{fig1}(c), where arrows indicate current amplitude and colors spin polarization. A magnetic field affects the QPC spin dependent transmission probability, enhancing one of the two charge currents and suppressing the other. The result is the flow of a net charge current $I_3$ in the QPC. The sign of $I_3$ reverses by reversing the magnetic field sign. Note that the net spin current remains constant in the three situations depicted in Fig.~\ref{fig1}(c).

By operating $\rm{QPC_3}$ as a floating probe, the charge current $I_3$ is constantly fixed to zero. In this case, the presence of a spin current $I^{(S)}_3$ in the QPC reflects itself in an antisymmetric component of the voltage behind it: $V_3(B) \ne V_3(-B)$. Remarkably, theory predicts that the zero-field derivative of the measured voltage, $\partial_B V_3$, is proportional to the spin current $I^{(S)}_3$ polarized along the applied magnetic field. The proportionality coefficient between the spin-to-charge signal $\partial_B V_3$ and $I^{(S)}_3$ is given by the QPC $g$ factor and its energy sensitivity $\hbar\omega$ measured at $N_3=0.5$ \cite{Stano2011}:

\begin{equation}
I_3^{(S)}=\frac{e^2}{h}\frac{2\hbar\omega}{\pi g \mu_B}\partial_BV_3.
\label{eq:spin_current1}
\end{equation}

More generally, for $N_3\leq1$ the presence of the spin current still reflects itself in a finite spin-to-charge signal whose amplitude is directly proportional to the detector normalized transconductance: $\partial_B V_3\propto\partial_VG_3/G_3$.  For $N_3=0.5$ it results in $\partial_V G_3/G_3=-\pi/\hbar\omega$. Details about the derivation of this proportionality and Eq.~(\ref{eq:spin_current1}) are reported in the Supplemental Material \cite{Note_supplemental}. Equation~(\ref{eq:spin_current1}) not only allows us to detect the presence of a spin current flowing in $\rm{QPC_3}$, but also to quantitatively measure and express it in units of ampere, giving the difference in currents carried by electrons with opposite polarization. Given the large SOI of our system, the measurement process requires only weak magnetic fields that do not affect the generated spin currents. We note that our approach is restricted to the linear response regime, and to terminal $3$ being a weak probe, i.e., $N_3\leq 1 \ll N_1,N_2$. The measurement protocol described here is independent of the spin current generation mechanism. In particular, the detection method can be used to measure spin currents of other origin.

\begin{figure}
\includegraphics[width=\columnwidth]{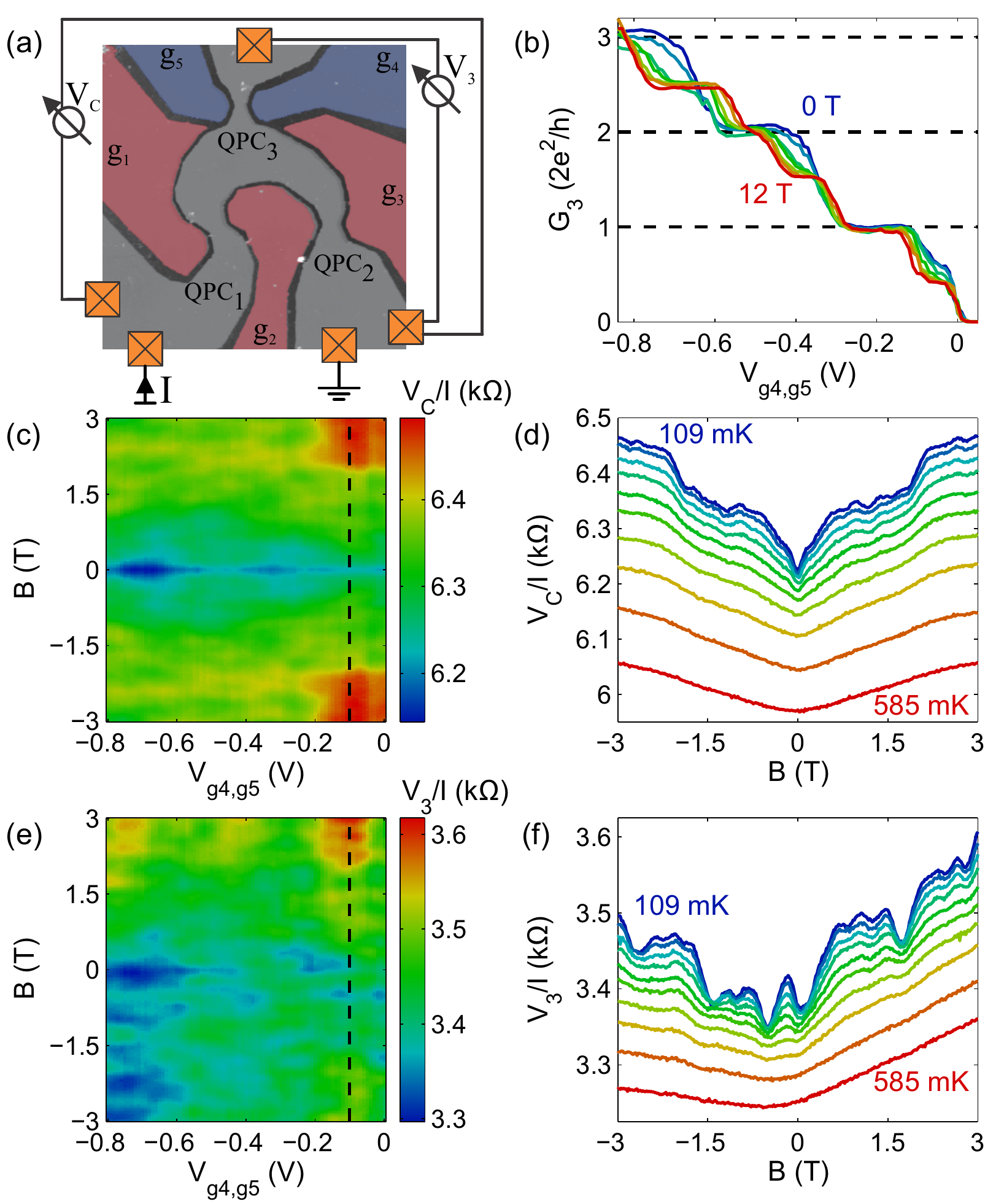}
\caption{(a) Atomic force micrograph of our sample, where dark lines indicate insulating trenches. The frame size is $5\times 5~\rm{\mu m^2}$. (b) $\rm{QPC_3}$ conductance as a function of the voltage applied to $\rm{g}_4$ and $\rm{g}_5$, for different values of magnetic field. (c) Cavity four-terminal resistance as a function of the voltage applied to $\rm{g}_4$ and $\rm{g}_5$ and magnetic field. (d) Linecut of (c) along the dashed line for different temperatures. (e) $V_3/I$ as a function of the voltage applied to $\rm{g}_4$ and $\rm{g}_5$ and magnetic field. (f) Linecut of (e) along the dashed line for different temperatures.}
\label{fig2}
\end{figure}

Motivated by the theory above, we study a three-terminal chaotic cavity embedded in a $p$-type GaAs two-dimensional hole gas (2DHG) with a strong Rashba SOI. Our sample, shown in Fig.~\ref{fig2}(a), lacks any spatial symmetry and consists of three leads and five in-plane gates, named $\rm{QPC}_{i}$ and $\rm{g}_{j}$ respectively. The gates $\rm{g}_4$ and $\rm{g}_5$, colored in blue, tune the conductance of $\rm{QPC_3}$ with little influence on the dot average conductance. The gates $\rm{g}_1$, $\rm{g}_2$ and $\rm{g}_3$, depicted in red, tune the conductance of $\rm{QPC}_1$ and $\rm{QPC}_2$, and also affect the dot shape. The lateral extent of the cavity is about $2~\rm{\mu m}$, the hole mean free path $l_e=4.8~\rm{\mu m}$ and the spin-orbit length $l_{SO}=36~\rm{nm}$ \cite{Note_supplemental}. Spin rotational symmetry is then completely broken and, with such a strong SOI, our cavity is in the so-called spin chaos regime \cite{Adagideli2010}. Unless differently stated, a charge current $I$ flows from terminal $1$ to terminal $2$, while terminal $3$ is connected to a high impedance voltage amplifier and is used to measure spin currents.

To measure the spin current in terminal $3$, we first extract the detector electric and magnetic sensitivity via a standard QPC characterization. Figure~\ref{fig2}(b) shows the detector conductance $G_3$ as a function of side gate voltage for different values of a magnetic field aligned with the $\rm{QPC_3}$ axis. The three well-developed plateaus visible at zero field split at finite field. The zero field slope and the finite field splitting give, respectively, $\partial_VG_3$ and the $g$ factor \cite{Note_supplemental}. 

After the detector calibration, $\rm{QPC_3}$ is operated as a voltage probe. The spin current measurement is performed running an AC current ($I=4~\rm{nA}$ unless stated otherwise) between terminals $1$ and $2$, and measuring the magnetic dependence of the voltages $V_C$ and $V_3$ as defined in Fig.~\ref{fig2}(a). The magnetic field is applied in-plane, to minimize its orbital effects, and aligned with the detector axis (unless stated otherwise). The finite zero field derivative $\partial_B V_3$, and its correlation with $\partial_VG_3/G_3$, indicates the presence of the spin current $I^{(S)}_3$.

\begin{figure}
\includegraphics[width=\columnwidth]{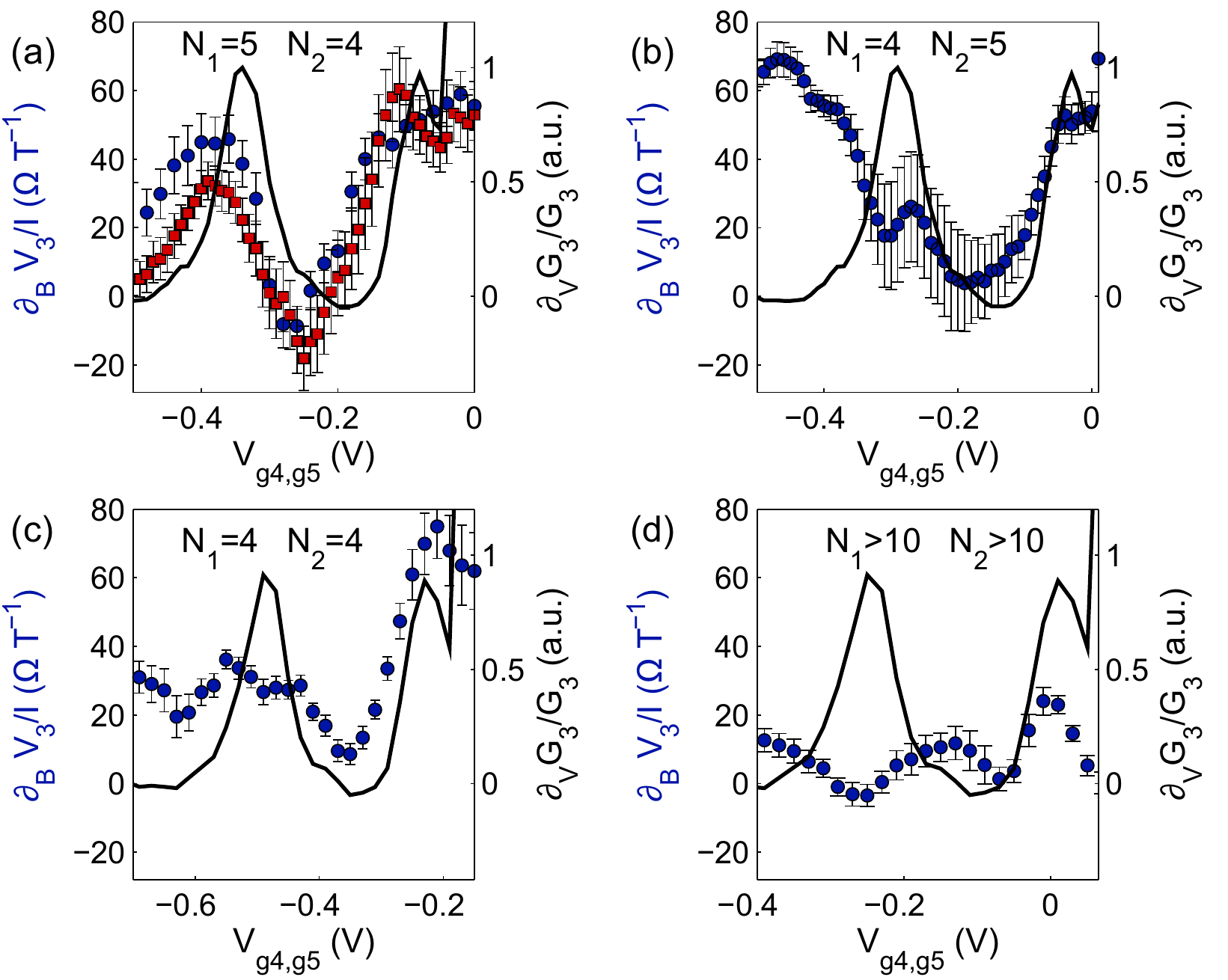}
\caption{(a)-(d) Comparison between $\partial_BV_3/I$ (markers) and $\partial_VG_3/G_3$ (solid line) as a function of side gate voltage for different number of modes in $\rm{QPC_1}$ and $\rm{QPC_2}$, as indicated in each subfigure. Dots and squares in (a) represent two identical measurements performed in different cooldowns \cite{comment_cooldown}.}
\label{fig3}
\end{figure}

Figures~\ref{fig2}(c) and (e) show the resistances $R_C=V_C/I$ and $V_3/I$ as a function of $B$ and the voltage applied to $\rm{g}_4$ and $\rm{g}_5$. Panels (d) and (f) show line cuts along the dashed lines of Fig.~\ref{fig2}(c) and (e), respectively, for different temperatures. These line cuts are taken for $N_3=0.5$. On top of a smooth background, higher frequency conductance fluctuations (CFs) appear at low temperatures. The cavity resistance $R_C$ is symmetric upon magnetic field reversal both in the slowly varying background and in the CFs, as expected from a two-terminal measurement in the linear regime \cite{Casimir1945}. $V_3$ is, on the contrary, strongly asymmetric. We first address the slowly varying background of $V_3(B)$. We will discuss the nature of the CFs below.

Large CFs do not allow us to measure meaningful voltage asymmetries at small magnetic fields. We therefore average them out by a linear regression of $V_3(B)$ in a magnetic field range between $-1~\rm{T}$ and $1~\rm{T}$. The chosen range is an optimal compromise between not sufficient CFs averaging at small fields and unwanted changes of the spin current at large fields. This interval includes at least six CFs, and we carefully checked that the averaged voltage asymmetry only weakly depends on the precise magnetic field range used for the analysis.

The detector voltage asymmetry extracted in this way is plotted in Fig.~\ref{fig3} (markers) together with the detector transconductance. We normalize the detector voltage by the constant cavity current $I$ and show the detector transconductance in arbitrary units. Panels (a)-(d) represent different cavity configurations, with different $N_1$ and $N_2$ as indicated in every subfigure. Despite the large error bars introduced by the CFs, in Figs.~\ref{fig3}(a), \ref{fig3}(b) and \ref{fig3}(c) we observe what theory anticipates: a correlation of the two quantities below the last detector conductance step (right-hand side of each subfigure) and disappearance of this correlation beyond the first plateau (left-hand side of each subfigure). This is the key observation from which we conclude that we measure a spin current. Where the detector is most energy sensitive, we observe useful signals with $\partial_BV_3/I\approx70~\rm{\Omega T^{-1}}$, with a typical background of $20~\rm{\Omega T^{-1}}$. The latter value was also typically observed for $N_1,N_2>6$, as shown in Fig.~\ref{fig3}(d), where the voltage asymmetry is uncorrelated with the transconductance. This is expected due to suppression of spin currents by energy averaging when many modes contribute to transport. In the following we give further evidence supporting the spin current origin of the observed voltage asymmetry.

\begin{figure}
\includegraphics[width=\columnwidth]{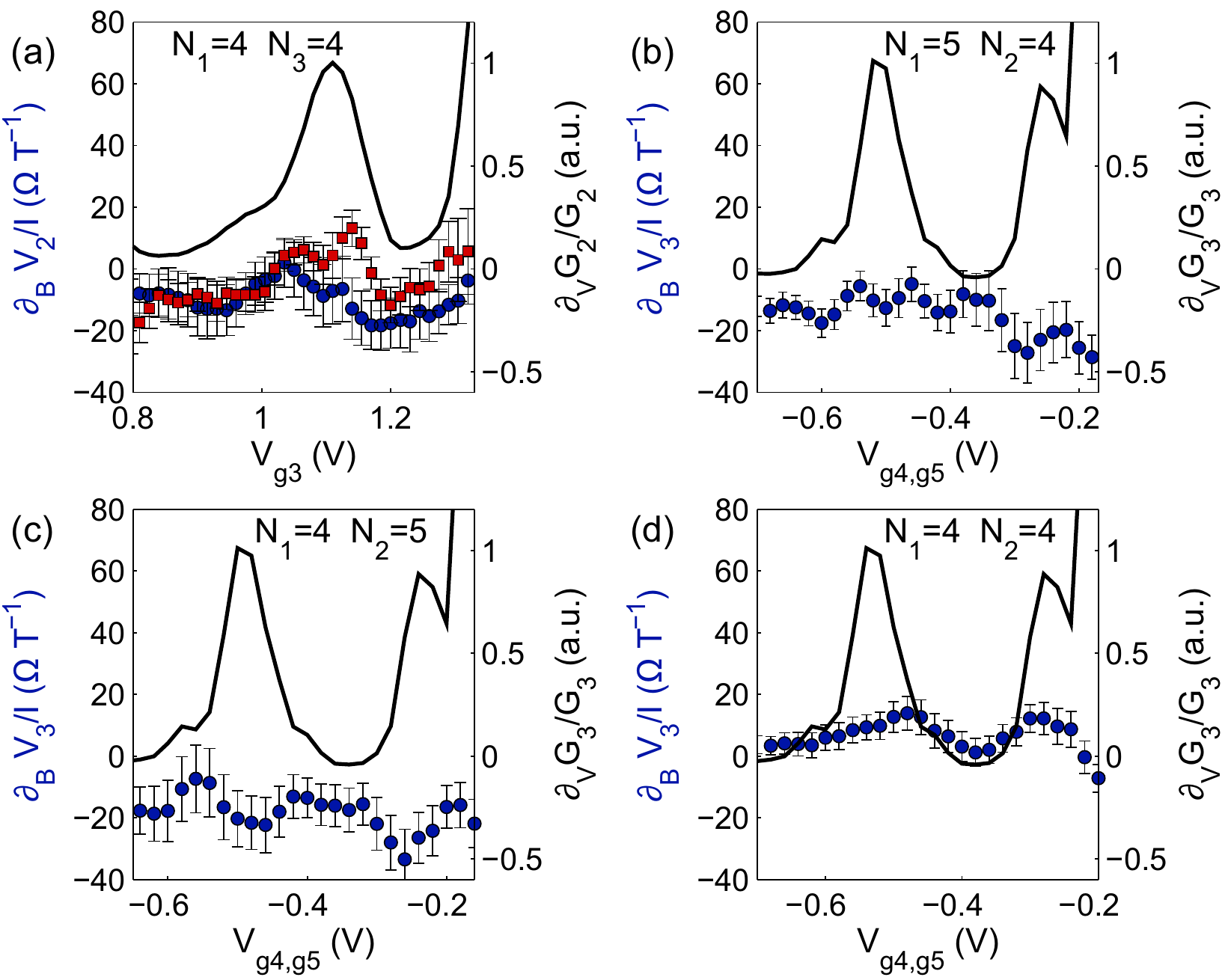}
\caption{(a) Comparison between $\partial_BV_2/I$ acquired with two different cavity shapes (dots and squares) both having $N_1=N_3=4$ and $\partial_VG_2/G_2$ (solid line) as a function of the voltage applied to $\rm{g}_3$. The electrical setup was modified to use $\rm{QPC_2}$ as detector. (b), (c) and (d) as in Fig.~\ref{fig3} but with the magnetic field aligned perpendicularly to $\rm{QPC_3}$.}
\label{fig4}
\end{figure}

To confirm the spin related nature of our signal, we exploit a key ingredient for the spin-to-charge conversion: the magnetic field sensitivity of the detector QPC. A detector with zero $g$ factor should result in a vanishing voltage asymmetry, regardless of the spin current intensity. We confirmed this prediction in the two ways shown in Fig.~\ref{fig4}. In Fig.~\ref{fig4}(a) we modified the measurement scheme such that the current flows from terminal $1$ to $3$, while terminal $2$ is used as the detector. The latter is characterized in Ref.~\cite{Nichele2014a}, and the first mode shows $g=0$. As expected, we observe a vanishing voltage asymmetry for $N_2\rightarrow 0$ for two different cavity shapes (dots and squares respectively). As yet an additional test, we kept the measurement scheme as in Fig.~\ref{fig2}(a), but rotated the sample by $90^\circ$ in the 2DHG plane, to have the magnetic field perpendicular to $\rm{QPC_3}$. Along this direction, the $g$ factor vanishes for all modes \cite{Note_supplemental}, which is a well-known anisotropy of $p$-type QPCs \cite{Srinivasan2012,Nichele2014a}. The latter is a particularly powerful approach as it leaves the spin current unaltered and only suppresses the spin-to-charge conversion efficiency. In Figs.~\ref{fig4}(b), \ref{fig4}(c) and \ref{fig4}(d) we show three of such measurements for the same mode configurations as in Figs.~\ref{fig3}(a), .~\ref{fig3}(b) and .~\ref{fig3}(c). In all the three cases, the voltage asymmetry is comparable to the background level of Fig.~\ref{fig3} and uncorrelated to the detector transconductance, proving the importance of a magnetic field sensitive measurement lead for observing an asymmetric voltage signal.

The spin-to-charge signal shown in Fig.~\ref{fig3} reflects a robust property of the system and is not linked to CFs. The CFs are phase coherent effects originating from electrostatic cavity shape distortion and magnetic flux penetration \cite{Beenakker1991}, or from a purely in-plane field as a consequence of the asymmetric and finite-width confinement potential \cite{Zumbuhl2004}. We carefully checked that our results do not depend on the CFs' pattern first by changing $\rm{g}_1$, $\rm{g}_2$ and $\rm{g}_3$ while keeping $N_1$ and $N_2$ constant, second by applying a strong voltage asymmetry on $\rm{g}_4$ and $\rm{g}_5$. In both cases, the CFs are completely changed without a significant modification of the voltage asymmetry extracted by averaging. Additionally, Fig.~\ref{fig3}(a) includes a measurement performed during a different cooldown (squares) \cite{comment_cooldown}. Despite the completely different CFs' fingerprint, an identical voltage asymmetry is obtained, proving the robustness of the measured effect.
As visible in Fig.~\ref{fig2}(f), coherent contributions are almost entirely suppressed at $T=530~\rm{mK}$, which is a standard temperature scale for the disappearance of coherent effects in quantum dots with few open channels \cite{Huibers1999}. The average signal is, on the other hand, more resistant to temperature increases because of its diffusionlike origin \cite{Adagideli2010}. We performed additional analysis of the temperature and charge current amplitude dependence of the spin-to-charge signal. These measurements, reported in the Supplemental Material \cite{Note_supplemental}, confirm the distinct nature of CFs and the spin-to-charge signal, as well as the fact that the spin-to-charge signal is a linear effect.

So far, we discussed the presence of a robust spin current in $\rm{QPC_3}$ visible from the slowing varying background of $\partial_BV_3/I$. As discussed in the context of the MSHE, CFs might also reflect the presence of mesoscopic spin CFs \cite{Ren2006,Bardarson2007,Nazarov2007,Krich2008}. Although the CFs occasionally show a finite zero field slope [see Fig.~\ref{fig2}(f)], it was not possible to univocally assign them to spin related or orbital effects, not considered in Ref.~\cite{Stano2011}. In particular, we could not test the $\rm{QPC_3}$ transmission dependence of $\partial_BV_3/I$ for single CFs due to the influence of $\rm{g}_4$ and $\rm{g}_5$ on the cavity shape.

We now evaluate the spin current amplitude for $N_3=0.5$. With the measured detector sensitivity $\hbar\omega=0.46~\rm{meV}$, its $g$~factor $g=0.27$ and the typical voltage asymmetry of $\partial_BV_3/I=60\rm{\Omega T^{-1}}$, Eq.~(\ref{eq:spin_current1}) gives $I_3^{(S)}=174~\rm{pA}$. We compare this value with theoretical predictions on geometrical correlation induced spin currents. The spin transmission of $\rm{QPC_3}$, calculated for a three-terminal cavity in the spin chaos regime, is \cite{Adagideli2010}:
\begin{equation}
\langle T_{13}^{(S)}\rangle=\mathcal{C}\frac{1+2\xi}{2l_{SO}k_{F}}\frac{N_1N_3}{N_1+N_2+N_3}\approx 0.137 \times \mathcal{C}.
\label{eq:spin_conductance}
\end{equation}
To evaluate this expression we used $\xi=1$, appropriate for a ballistic dot, $N_1=N_2=4$, $N_3=0.5$. $\mathcal{C}$ is a system specific prefactor, of order unity. Neglecting spin flips caused by $\rm{QPC_3}$ itself, the expected spin current is
\begin{equation}
I_3^{(S)}\approx \frac{e^2}{h} \langle T_{13}^{(S)}\rangle (V_1-V_2)=\mathcal{C}\times 134~\rm{pA} 
\end{equation}
for a charge current of $4~\rm{nA}$. The very good agreement with our measurement further supports the interpretation that our signal goes beyond a mere spin current detection, but provides a quantitatively reliable magnitude. As shown by Eq.~\ref{eq:spin_conductance}, the spin conductance depends on $N_3$, allowing larger spin currents to be generated by opening $\rm{QPC}_3$ further. However, since the detection scheme requires $N_3=0.5$, we could not probe this scenario.

Similarly to bulk materials, the spin-to-charge conversion efficiency of the cavity can be expressed via the spin Hall angle $\Theta$, defined as the ratio between spin and charge current densities. In our case we get $\Theta=(I^{(S)}_3 /N_3)/(I/N_2)\approx34\%$, independent on $N_3$ for $N_3\ll N_1,N_2$. This efficiency is substantially higher than any reported for semiconductors \cite{Jungwirth2012}, making this system interesting for future semiconductor spintronics applications.

\setcounter{equation}{0}
\renewcommand{\theequation}{S.\arabic{equation}}
\setcounter{figure}{0}
\renewcommand{\thefigure}{S.\arabic{figure}}

\section{Supplementary Material}
In this Supplemental Material section we provide additional information on the wafer structure used for the experiment and on the techniques adopted to measure the sample and characterize the detector QPC. We furthermore show the current and temperature dependence of the spin-to-charge signal discussed in the main text. We provide an analytical treatment of the spin-to-charge conversion effect directly applicable to our experiment and we discuss its dependence on the detector QPC conductance.


\maketitle


\subsection{Materials and Methods}
The two-dimensional hole gas (2DHG) in use consists of a $15~\rm{nm}$ GaAs quantum well placed $45~\rm{nm}$ below the surface. The sample is grown along the [001] direction and remotely doped with carbon. The wafer has been extensively characterized by magnetotransport measurements in Ref.~\cite{Nichele2014}. The total hole density is $n=3.0\times 10^{15}~\rm{m^{-2}}$ and the strong Rashba spin-orbit interaction (SOI) results in a splitting of the dispersion relation in the two subbands with different total angular momentum projection along the growth direction. The densities of the two spin-orbit split subbands have been derived from the periodicity of the Shubnikov-de Haas oscillations as $n_1=1.05 \times 10^{15}~\rm{m^{-2}}$ and $n_2=1.95 \times 10^{15}~\rm{m^{-2}}$ resulting in a cubic Rashba parameter $\beta=1.4\times 10^{-28}\rm{eVm^3}$ \cite{Winkler2003}. The spin-orbit length, defined as the length scale over which the electron spin rotates by $2\pi$ is calculated, as in Ref.~\cite{Adagideli2010}, as $l_{SO}=v_F\tau_{SO}=(\hbar k_F/m^*)(2\hbar/\Delta_{SO})$. $\Delta_{SO}=2\beta_Rk_F^3$ is the spin-orbit energy splitting, $m^*$ the hole effective mass and $k_F$ the Fermi wavevector. Because of the large difference in $m^*$ and $k_F$ between the two spin-orbit split subbands \cite{Nichele2014}, we use their density-averaged values $m^*=0.71m_e$ and $k_F=1.43\times10^8~\rm{m^{-1}}$.

Two nominally identical cavities were defined by electron beam lithography and wet etching (see Fig. 1(a) of the main text). The samples were measured in a dilution refrigerator with a base temperature of $110~\rm{mK}$ using standard low-frequency lock-in techniques. The tilt angle between 2DHG and magnetic field could be tuned \textit{in-situ} with an accuracy of less than $0.05^\circ$. The asymmetry in $V_3(B)$ and the conductance fluctuations (CFs) visible in Fig.~2(f) of the main text are genuine effects due to the in-plane magnetic field only. Tilting the 2DHG angle with respect to the external magnetic field between $-0.25^\circ$ and $0.25^\circ$ leaves $V_3(B)$ unaffected, proving that an eventual residual out-of-plane component of the magnetic field is negligible for the effects under discussion. The voltage measurements are performed in a longitudinal configuration, so that an out-of-plane magnetic field would not result in the appearance of a Hall slope. Changing the orientation of the device with respect to the in-plane component of the magnetic field required warming up the sample and manually changing its bonding configuration. The electronic properties of the sample and the characteristics of the QPC remained largely unchanged by this carefully done procedure. The two devices showed quantitatively comparable behavior, reproducible over multiple cool-downs. In the main text we show data from a single sample, characterized by a larger tunability of the three QPCs used as leads.

\subsection{Characterization of the detector QPC}
In the following we describe in more detail the characterization measurements performed on $\rm{QPC_3}$. The data presented allows the extraction of the QPC $g$-factor and energy sensitivity, used to quantify the spin current intensity from the spin-to-charge signal. Furthermore, we show the existence of a pronounced Zeeman splitting anisotropy that allowed us to obtain the data presented in Fig.~4 (b), (c) and (d) of the main text. A similar procedure has been discussed in greater detail in Ref.~\cite{Srinivasan2012,Nichele2014a}.

\begin{figure}
\includegraphics[width=\columnwidth]{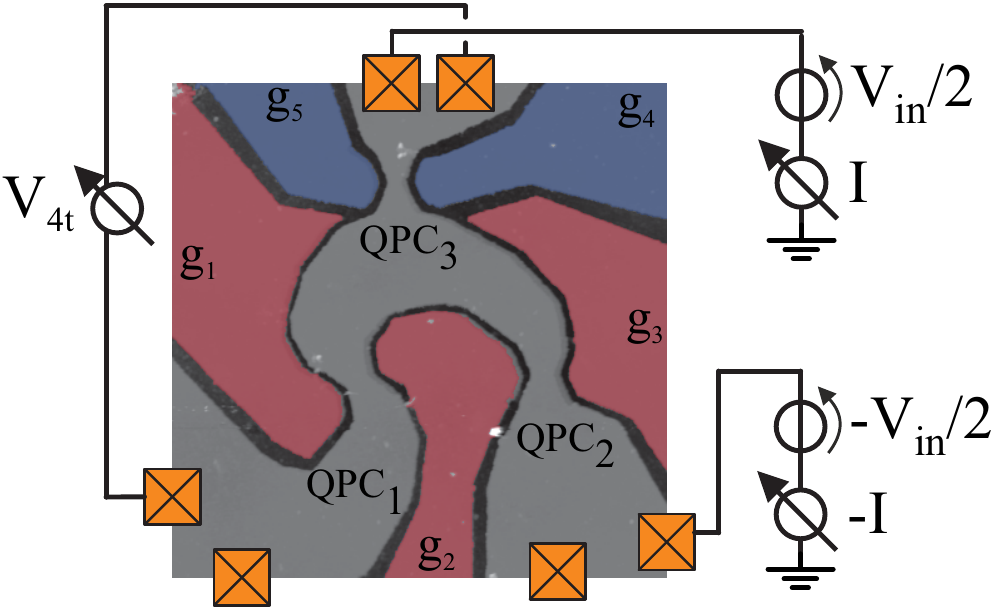}
\caption{False color atomic force micrograph of the sample under study, together with a schematic of the electrical setup used to characterize $\rm{QPC_3}$.}
\label{sup1}
\end{figure}

\begin{figure*}
\includegraphics[width=16cm]{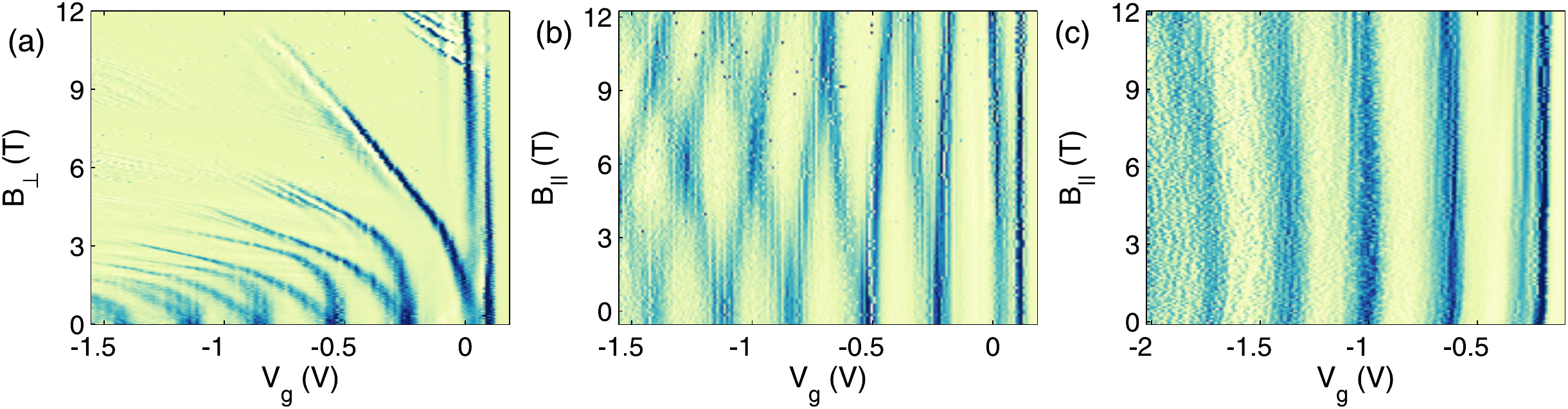}
\caption{$QPC_3$ transconductance (numerical derivative of the linear conductance with respect to the gate voltage axis) as a function of gate voltage and magnetic field for different field orientations. Dark lines indicate high transconductance, bright regions indicate low transconductance. (a) Magnetic field applied perpendicular to the plane of the 2DHG. (b) Magnetic field applied in the plane of the 2DHG and along the QPC axis. (c)  Magnetic field applied in the plane of the 2DHG and perpendicular to the QPC axis.}
\label{sup2}
\end{figure*}

Fig.~\ref{sup1} shows the electrical setup used for the characterization of $\rm{QPC_3}$. A low-frequency AC bias $V_{in}=15\rm{\mu V}$ was symmetrically applied to the cavity leads $2$ and $3$ and the current flowing in the structure was measured as a function of the voltage applied to $g_4$ and $g_5$. At the same time, the voltage drop between terminal $1$ and $3$ was recorded, allowing to extract a four terminal QPC conductance that does not depend on $\rm{QPC_1}$ and $\rm{QPC_2}$. During the characterization of $\rm{QPC_3}$, $\rm{QPC_1}$ and $\rm{QPC_2}$ were set to a very high conductance by negatively biasing $g_1$, $g_2$ and $g_3$. We carefully checked that the precise voltages applied to  $g_1$, $g_2$ and $g_3$ have no effects on the presented results.

Via finite bias measurements, we extracted the voltage dependent lever arm of $g_4$ and $g_5$. The lever arm allows one to convert the gate voltage axis into energy and extract quantities such as the $g$-factor and the energy sensitivity $\hbar\omega$ from a conductance measurement. For the experiment under consideration, the determination of the lever arm is irrelevant if both $\hbar\omega$ and the $g$-factor are extracted from the same conductance plot and in a narrow gate voltage range, as we do here. In fact, as shown by Eq.~(1) of the main text, the spin-to-charge conversion amplitude is only given by the ratio $\hbar\omega/g$ that does not depend on the gate lever arm.

Close to a conductance $G_3=e^2/h$, the relevant regime for the effects presented in the main text, the lever arm was measured to be $4~\rm{meV V^{-1}}$. We determined $\hbar\omega$, i.e. the curvature of the harmonic potential, by fitting a saddle point model \cite{Buttiker1990} to the QPC conductance \cite{Rossler2011}. Since we do know the lever arm in this case, we can convert the curvature into energy units by fitting the equation:
\begin{equation}
G(E)=\frac{2e^2}{h}N\left(\frac{1}{1+\exp\left(2\pi(E-E_0)/(\hbar\omega)\right)}\right),
\end{equation}
with the fitting parameters being a constant energy shift $E_0$ and the potential curvature $\hbar\omega$. $N$ is the mode number of the plateau under consideration, in this case $N=1$.

The QPC is characterized by a strong $g$-factor anisotropy, typical for QPCs embedded in 2DHGs grown along the [001] crystallographic direction \cite{Srinivasan2012,Nichele2014a}. The QPC transconductance (numerical derivative with respect to gate voltage) is shown in Fig.~\ref{sup2} for three different magnetic field orientations. A finite spin splitting is present when the magnetic field is oriented perpendicularly to the plane of the 2DHG (Fig.~\ref{sup2}(a)) or in the plane of the 2DHG and aligned along the QPC axis (Fig.~\ref{sup2}(c)). No spin splitting up to $12~\rm{T}$ is visible when the magnetic field is aligned in the plane of the 2DHG and perpendicular to the QPC axis (Fig.~\ref{sup2}(b)). In order to use the QPC to detect a spin current via the spin-to-charge conversion scheme, it is necessary to have a finite $g$-factor (see Eq.~(1) of the main text). In the present case, this is possible by performing the experiment with the magnetic field oriented as in Fig.~\ref{sup2}(b). The $g$-factor is obtained from Fig.~\ref{sup2}(b) by tracking the separation in gate voltage between spin split plateaus as a function of an in-plane magnetic field. The gate voltage separation is then converted into Zeeman energy using the gate dependent lever arm. For the first mode we find $g=0.27$. In order to suppress the spin-to-charge conversion efficiency, the magnetic field must be turned by $90^\circ$ in the plane of the 2DHG, obtained in the situation shown in Fig.~\ref{sup2}(c). In this case all the modes are characterized by $g=0$.

As visible in Fig.~\ref{sup2}(b), the levels splitting as a function of an in-plane magnetic field is not symmetric in gate voltage. For each pair of spin-split levels, the left branch rises in energy (more negative values of gate voltage) faster then the right branch. The anomalous magnetic field dependence of the QPC levels in an in-plane magnetic field was recently studied in Ref.~\cite{Nichele2014a} and found to be caused by a quadratic SOI peculiar for hole gases. This anomalous spin-splitting makes the points of strongest magnetic field dependence of $G_3$ not to correspond to the points of strongest gate-voltage dependence, as assumed in Ref.~\cite{Stano2011}, but to be slightly shifted to more negative gate voltage. This effect might explain the small horizontal offset between $\partial_BV_3/I$ and $\partial_VG_3/G_3$ seen in Fig.~3(a), (b) and (c) of the main text.

\subsection{Current and temperature dependence of the spin-to-charge signal}

\begin{figure}
\includegraphics[width=\columnwidth]{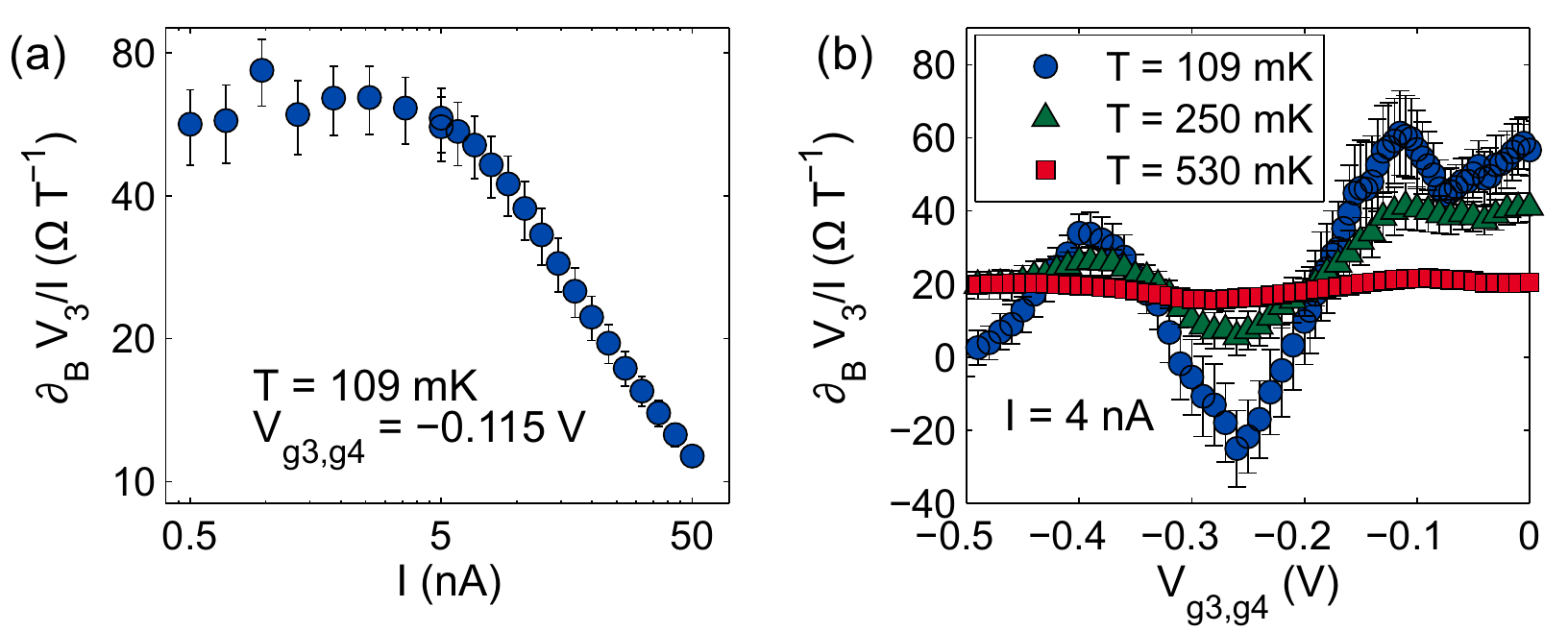}
\caption{(a) $\partial_BV_3/I$ as in Fig.~3(a) of the main text and $N_3=0.5$ for different current intensities. (b) $\partial_BV_3/I$ as in Fig.~3(a) of the main text for different temperatures.}
\label{supTdep}
\end{figure}

The spin current detection method we employ is limited to the linear transport regime~\cite{Stano2012}, estimated to break-down in our samples for currents of a few nA. In the linear regime, $\partial_BV_3/I$ should be independent of $I$. We plot this current dependence in Fig.~\ref{supTdep}(a) for the same configuration as Fig.~2(a) of the main text and $N_3=0.5$. Remarkably, the value of $\partial_BV_3/I$ is independent of $I$ for $I<5~\rm{nA}$, confirming the hypothesis that the slope in $V_3(B)$ is a linear effect. For currents higher than $5~\rm{nA}$, in addition to sample heating, we enter the non-linear transport regime where the detector voltage asymmetry and transconductance are not necessarily linked.

Fig~\ref{supTdep}(b) shows the temperature dependence of the extracted $\partial_BV_3/I$ as in Fig.~3(a) of the main text, for three different temperatures. At $T=530~\rm{mK}$ we still recover the same gate dependence of $\partial_BV_3$ as shown in Fig.~3(a)of the main text, but with a reduction in peak-to-peak height by a factor of $15$. The signal eventually disappears entirely at higher temperature because of the ensuing energy-averaging, affecting also the geometric correlation corrections. Furthermore, for $T>600~\rm{mK}$, $G_3$ loses the step-like behavior (hence its energy sensitivity), becoming a smooth function of the gate voltage.

\subsection{Spin-to-charge conversion}
We review here the theoretical basis for the spin-to-charge conversion effect in a three-terminal mesoscopic cavity as the one depicted in Fig.~1(a) of the main text. We extend the theory presented in Ref.~\cite{Stano2011} to prove the proportionality between $\partial_BV_3/I$ and $\partial_VG_3/G_3$.

The cavity under consideration has three contacts (labeled 1,2 and 3), and each of them is at a voltage $V_i$ and carries $N_i$ spin degenerate modes. We consider the situation where $N_3\leq 1$ and $N_1,N_2\gg 1$. By applying a voltage bias $V_2-V_1$ between leads $1$ and $2$, a charge current $I$ will flow in the cavity. If contact $3$ is grounded, a charge current $I_3^{(0)}$ will flow in it. In the following we are interested in the situation in which no charge current flows in contact $3$ at zero magnetic field. This situation can be achieved either by applying a voltage $V_3$ that sets $I_3^{(0)}$ to zero at zero magnetic field, or by leaving it floating and connected to a volt meter that measures $V_3$. In the first case, the application of a magnetic field will make the current vary from zero, in the second case the current will always remain zero and the measured $V_3$ will vary. As we will show in the following, the zero field spin current $I_3^{(S)}$ in contact $3$ is directly proportional to the zero-field derivative of $I_3$, or $V_3$, with respect to the in-plane magnetic field:

\begin{align}
I_{3}^{(S)}&=\frac{\hbar\omega}{\pi\mu}\partial_B I_{3}^{(0)}\arrowvert_{B=0} \label{eq:C9_results1},\\
I_{3}^{(S)}&=\frac{\hbar\omega}{\pi\mu}\frac{e^2}{h}\left(2-T_{33}^{(0)}\right)\partial_B V_{3}\arrowvert_{B=0},
\label{eq:C9_results2}
\end{align}
where $\hbar\omega$ is the magnetic field energy sensitivity, $\mu=g\mu_B/2$ is the electron magnetic moment and $(2-T_{33}^{(0)})$ is the charge transmission coefficient of contact $3$. These QPC parameters are easily measured experimentally.

In the following theoretical treatment we do not include the presence of any orbital effect. The latter assumption is well justified if no out-of-plane magnetic field is applied. An in-plane field can give rise to other orbital effects \cite{Zumbuhl2004} which will not be accounted for in this section, assuming that these are small enough. We will always consider that the voltages applied to the system are within the linear response regime (i.e. small compared to other energy scales). The generic current $I_i^{(\alpha)}$ at a contact $i$ can be calculated using Landauer-B\"uttiker formalism, resulting in:

\begin{equation}
I_i^{\alpha}=\sum\limits_j{\left(2N_j\delta_{ij}\delta_{\alpha 0}-T_{ij}^{(\alpha)}\right)V_j},
\label{eq:c8_LB1}
\end{equation}
where $i=1,2,3$ denotes the leads and $\alpha=0,x,y,z$ denotes the spin polarization of the current. The generic transmission coefficient is:

\begin{equation}
T_{ij}^{(\alpha)}=\sum\limits_{m\in i, n\in j}{Tr\left( t_{mn}^\dagger\sigma^{(\alpha)}t_{mn}\right)},
\label{eq:c8_LB2}
\end{equation}
where $\sigma^{(\alpha)}$ are spin matrices, with $\sigma^{(0)}$ the identity matrix. The $2\times 2$ matrices $t_{mn}$ indicate the probability of an electron entering the cavity from the $n$-th mode of QPC $j$ to exit the cavity from the $m$-th mode of QPC $i$, their elements $t_{m,n}^{\sigma\sigma'}$ take into account spin flipping. It can be shown that the transmission coefficients for charge and spin are:

\begin{eqnarray}
\tau_{ij}^{(0)}=\sum\limits_{\sigma\sigma'}{\tau_{ij}^{\sigma\sigma'}} \label{eq:c9_LB3a}\\,
\tau_{ij}^{(S)}=\sum\limits_{\sigma\sigma'}{\sigma\tau_{ij}^{\sigma\sigma'}}.
\label{eq:c9_LB3b}
\end{eqnarray}
The transmission probabilities from contact $1$ or $2$ to contact $3$ are assumed to take the form:
\begin{equation}
T_{3i}^{\sigma\sigma'}(B)=\tau_{3i}^{\sigma\sigma'}(B)\Gamma\left(E_F-\sigma\mu B\right),
\label{eq:c9_LB4}
\end{equation}
hence they can be separated into a spin dependent part and an energy dependent part. The spin affects the second term only via the Zeeman energy. In Eq.~(\ref{eq:c9_LB4}) it was assumed that the QPC has high energy sensitivity, hence $\Gamma\left(E_F-\sigma\mu B\right)$ varies faster than $\tau_{3i}^{\sigma\sigma'}(B)$ with $B$. $\tau_{3i}^{\sigma\sigma'}(B)$ are phenomenological parameters, describing the spin transmission of the QPC when it is fully open.  Eq.~(\ref{eq:c9_LB4}) is valid only in the limit when an electron reflected back in the cavity from contact $3$ has a negligible probability to come back to contact $3$ again. This limit is achieved when $N_1,N_2\gg N_3$.

Using Landauer-B\"uttiker expressions for charge and spin current through contact $3$ one has:
\begin{eqnarray}
I_3^{(0)}=\frac{e^2}{\hbar}\left( T_{31}^{(0)}(V_3-V_1)+T_{32}^{(0)}(V_3-V_2) \right),\\
I_3^{(S)}=\frac{e^2}{\hbar}\left( T_{31}^{(S)}(V_3-V_1)+T_{32}^{(S)}(V_3-V_2) \right).
\label{eq:c9_LB5}
\end{eqnarray}
Imposing $I_3(0)=0$ allows to find an analytical form for the spin current:

\begin{align}
V_3 & =\left(T_{31}^{(0)}V_1T_{32}^{(0)}V_2\right)/\left(T_{31}^{(0)}+T_{32}^{(0)}\right),\\
I_3^{(S)} & =\frac{e^2}{\hbar}\left( T_{31}^{(S)}(V_3-V_1)+T_{32}^{(S)}(V_3-V_2) \right).
\label{eq:c9_LB6}
\end{align}
Equations (\ref{eq:C9_results1}) and (\ref{eq:C9_results2}) are obtained by evaluating $\partial_B {I_{3}}_{(0)\arrowvert_{B=0}}$ with constant $V_3$ and $\partial_B V_3\arrowvert_{B=0}$ for $I_3^{(0)}=0$, respectively. For obtaining a simpler analytical form of these expressions, we will make an assumption of the specific form of the QPC transmission $\Gamma$, though the results of this section do not qualitatively depend on this choice. We choose  the saddle point potential model, which gives the QPC energy dependent transmission probability as \cite{Buttiker1990}:

\begin{equation}
\Gamma(E_F,V_g,B)=\frac{1}{ 1 + \exp{ \left(-2\pi\left( E_F - \alpha V_g -\sigma\mu B\right) /{ \hbar \omega} \right)}}.
\label{eq:c9_QPC1}
\end{equation}
The partial derivative of the QPC transmission with respect to magnetic field is:
\begin{equation}
\partial_B\Gamma(E_F,V_g,B)=\sigma\mu\partial_{V_g}\Gamma(E_F,V_g,B).
\label{eq:c9_QPC2}
\end{equation}
This allows us to write the magnetic field derivative of the charge transmission coefficients $T_{3i}^{(0)}$ in terms of spin transmission coefficients $T_{3i}^{(S)}$:
\begin{align}
\partial_B T_{3i}^{(0)} & =\partial_B \sum \limits_{\sigma\sigma'}{T_{ij}^{\sigma\sigma'}\arrowvert_{B=0}}\\
& =\sum \limits_{\sigma\sigma'}{\tau_{ij}^{\sigma\sigma'}\partial_B\Gamma(E_F,V_g,0)\arrowvert_{B=0}}\\
& =\sum \limits_{\sigma\sigma'}{\tau_{ij}^{\sigma\sigma'}\sigma\mu\partial_{V_g}\Gamma(E_F,V_g,0)\arrowvert_{B=0}}\\
& =\sum \limits_{\sigma\sigma'}{\tau_{ij}^{\sigma\sigma'}\sigma\mu\Gamma(E_F,V_g,0)\frac{\partial_{V_g}\Gamma(E_F,V_g,0)\arrowvert_{B=0}}{\Gamma(E_F,V_g,0)}}\\
& =\mu T_{3i}^{(S)}\frac{\partial_{V_g}\Gamma(E_F,V_g,0)\arrowvert_{B=0}}{\Gamma(E_F,V_g,0)}.
\label{eq:c9_partialI}
\end{align}
In the middle of the first slope in the QPC conductance, the energy sensitivity is maximal and it gives ($V_g=E_F$):
\begin{equation}
\frac{\partial_{V_g}\Gamma(E_F,V_g,0)\arrowvert_{B=0}}{\Gamma(E_F,V_g,0)}=-\frac{\pi}{\hbar\omega}.
\label{eq:c9_QPC3}
\end{equation}
With this, the zero-field derivative of the charge current is ($V_1$, $V_2$, $V_3$ are constant):
\begin{widetext}
\begin{align}
\partial_B I_3^{(0)}\arrowvert_{B=0} & =-\frac{e^2}{\hbar}\left( \partial_B T_{31}^{(0)}(V_3-V_1)+\partial_B T_{32}^{(0)}(V3-V_2) \right)\\
& = -\frac{e^2}{\hbar} \left( \mu \frac{\partial_{V_g}\Gamma(E_F,V_g,0)\arrowvert_{B=0}}{\Gamma(E_F,V_g,0)}\right)\left(T_{31}^{(S)}(V_3-V_1)+T_{32}^{(S)}(V_3-V_2)\right)\\
& = -\mu \frac{\partial_{V_g}\Gamma(E_F,V_g,0)\arrowvert_{B=0}}{\Gamma(E_F,V_g,0)} I_3^{(S)}\\
& = \frac{\pi\mu}{\hbar\omega}I_3^{(S)}.
\label{eq:c9_LB7_last}
\end{align}

Finally, Eq~(\ref{eq:C9_results1}) is obtained solving Eq.~(\ref{eq:c9_LB7_last}) for $I_3^{(S)}$.
Similarly, the zero field derivative of $V_3$ is: 
\begin{equation}
\partial_BV_3\arrowvert_{B=0} =-\mu\frac{\partial_{V_g}\Gamma(E_F,V_g,0)\arrowvert_{B=0}}{\Gamma(E_F,V_g,0)} \frac{1}{2-T_{33}^{(0)}} \frac{h}{e^2}I_3^{(S)}.
\label{eq:c9_LB8}
\end{equation}
\end{widetext}
In the point of highest energy sensitivity we have:
\begin{equation}
\partial_BV_3\arrowvert_{B=0}  = \frac{h}{e^2} \frac{\pi\mu}{\hbar\omega}\frac{1}{2-T_{33}^{(0)}}I_3^{(S)},
\label{eq:c9_LB9}
\end{equation}
which results in Eq.~(\ref{eq:C9_results2}) upon solving for $I_3^{(S)}$.

It is interesting, in the light of our experimental results, to investigate the behavior of $\partial_B V_3\arrowvert_{B=0}$ around the point of highest energy sensitivity. $2-T_{33}^{(0)}$ is the exact expression for the charge current transmission coefficient of contact $3$. It can be approximated as (see Eq.~(\ref{eq:c9_LB4})):
\begin{widetext}
\begin{equation}
2-T_{33}^{(0)}=T_{31}^{(0)}+T_{32}^{(0)}=\sum \limits_{\sigma\sigma'}{\left(\tau_{31}^{\sigma\sigma'}+\tau_{32}^{\sigma\sigma'}\right)}\Gamma(E_F,Vg,B).
\end{equation}
We can apply the approximation from Eq.~(\ref{eq:c9_LB4}) to Eq.~(\ref{eq:c9_LB5}), finding a relation between $I_3^{(S)}$ and $\Gamma$:
\begin{equation}
I_3^{(S)}=-\frac{e^2}{h}\left( \sum \limits_{\sigma\sigma'} {\tau_{31}^{\sigma\sigma'}}(V_1-V_3)+\sum \limits_{\sigma\sigma'} {\tau_{32}^{\sigma\sigma'}}(V_2-V_3)\right)\Gamma(E_F,V_g,B).
\end{equation}
\end{widetext}
Combining the last three equations we get the expression:
\begin{equation}
\partial_B V_3 \arrowvert_{B=0}=\mu C\frac{\partial_{V_g}\Gamma(E_F,V_g,0)\arrowvert_{B=0}}{\Gamma(E_F,V_g,0)},
\end{equation}
where $C$ is a prefactor containing the voltages $V_i$ and the coefficients $\tau_{3i}^{\sigma\sigma'}$. It is assumed to be constant with respect to gate voltage. The results obtained here can be summarized with the following proportionality relation:
\begin{equation}
\label{eq:c9_partialV3}
\partial_B V_3 \arrowvert_{B=0} \propto \frac{\partial_{V_g}\Gamma(E_F,V_g,0)\arrowvert_{B=0}}{\Gamma(E_F,V_g,0)}.
\end{equation}
It is worth reminding that the last proportionality is valid in the limit $N_1,N_2\gg N_3$ and $N_3\leq 1$ and the coefficients $\tau_{3i}^{\sigma\sigma'}$ are supposed to weakly depend on magnetic field.

In the absence of SOI, reversing the magnetic field direction reverses the sign of the spin polarization $S$. In this case $\tau_{3i}^{(0)}(B)=\tau_{3i}^{(0)}(-B)$ and $\tau_{3i}^{(S)}(B)=-\tau_{3i}^{(S)}(-B)$ (see Eq.~(\ref{eq:c9_LB3a}) and (\ref{eq:c9_LB3b})). Since $\Gamma(E_F,V_g,B)$ is an even function of $B$, it results that $T_{3i}^{\sigma\sigma'}(B)=-T_{3i}^{\sigma\sigma'}(-B)$, and $I_3^{S}(0)=0$ (see Eq.~(\ref{eq:c9_LB6})): in the absence of SOI, both $\partial_B I_{3}^{(0)}$ and $\partial_B V_{3}$ vanish.

\begin{acknowledgments}
We acknowledge Christian Gerl for growing the wafer structure. The authors wish to thank the Swiss National Science Foundation via NCCR QSIT ``Quantum Science and Technology" for financial support.
\end{acknowledgments}

\bibliography{Bibliography}

\begin{thebibliography}{27}
\bibitem [{com()}]{comment_cooldown}%
  \BibitemOpen
  \href@noop {} {}\bibinfo {note} {For a direct comparison, we horizontally
  shifted the red curve to account for the different pinch-off voltage of
  $\rm{QPC_3}$ in this cool-down.}\BibitemShut {Stop}%
\end{thebibliography}
\end{document}